\documentclass[a4paper,11pt]{amsart}
\begin{document}
\hyphenation{ge-ne-ra-tion spe-ci-fying exact}
\title[] {{\bf  An immediate proof\\of the non-existence of GW's}}
\author[]{Angelo Loinger}
\address{Dipartimento di Fisica, Universit\`a di Milano, Via
Celoria, 16 - 20133 Milano (Italy)}
\email{angelo.loinger@mi.infn.it}
\thanks{To be published on \emph{Spacetime \& Substance.}}

\begin{abstract}
In general relativity (GR) no observer is physically privileged.
As a strict consequence, it can be shown that the physical
generation of gravitational waves (GW's) is quite impossible.
\end{abstract}

\maketitle

\vskip1.20cm
\textbf{1}. -  As it is well known, the notion of GW came forth as
a by-product of the \textbf{linear} approximation of GR \cite{1}.
Now, this approximation -- which \emph{resembles} Maxwell e.m.
theory -- is fully \textbf{inadequate} to a proper study of the
hypothetic GW's (see \cite{2}, \cite{3}). On the other hand, in
the \textbf{exact} (non-approximate) formulation of GR \textbf{no}
``mechanism'' exists in reality for the \emph{physical generation}
of the GW's, as it can be proved \cite{3}. The undulatory
solutions of Einstein field equations have a mere \emph{formal}
character.

\par I give here \textbf{another} proof of the real non-existence of
\emph{physical} GW's, which is so straightforward that even the
physicist in the street will understand it.

\vskip0.50cm
\textbf{2}. -  There is a \textbf{radical} difference between
Maxwell e.m. theory and Einstein general relativity: Maxwell
theory is Lorentz invariant, the set of the inertial observers is
\emph{physically} privileged -- Einstein theory has an invariant
character with respect to \emph{all} transformations of general
co-ordinates, and \emph{no} observer is \emph{physically}
privileged \cite{4}. \emph{All} observers are on an equal physical
footing just like the \emph{inertial} observers in Maxwell theory.
(Strictly speaking, in GR the concept ``observer'' requires a
particular specification \cite{5}).

\vskip0.50cm
\textbf{3}. -  An electric charge $C$ which is \emph{at rest} for
a given \emph{inertial} observer $I_{0}$ cannot emit e.m. waves.
Any inertial observer $I$ for whom $C$ is in motion does not
possess physical privileges with respect to $I_{0}$. Accordingly,
both observer $I_{0}$ and observer $I$ do not register any e.m.
wave sent forth by $C$.

\par In GR the expression \emph{at rest} must be defined precisely
every time, \emph{specifying the interested spacetime manifold},
because the co-ordinates are mere ``labels'' of point events
\cite{5}. Let us consider for instance the Einsteinian
gravitational field generated by a homogeneous sphere of an
incompressible fluid as it was investigated by Schwarzschild
\cite{6}. In Schwarzschild's system of co-ordinates $S_{0}$ the
sphere is at rest, and no GW is emitted. Now, any observer $S$ --
very far, in particular, from observer $S_{0}$ --, for whom the
sphere of fluid is in motion does not possess any physical
privilege with respect to $S_{0}$. Accordingly, both observer
$S_{0}$ and observer $S$ do not register any GW sent forth by the
sphere of fluid. It is evident that this argument can be extended
to any celestial body $B$, which can be considered at rest for a
given observer $S_{0}$ and in motion for any observer $S$, who is
very far, in particular, from $S_{0}$. \emph{Q.e.d.} --

\vskip1.00cm
\begin{center}
\noindent \emph{\textbf{APPENDIX}}
 \vskip0.50cm\nopagebreak
\end{center}

$\alpha$) I have repeatedly emphasized in my papers that the
notion of GW was originally (1916-1918) a by-product of the
\textbf{linearized} version of the exact (nonlinear) formulation
of GR. This circumstance has unduly influenced almost all the
subsequent researches (both theoretical and experimental)
concerning the GW's. The great majority of the physicists have
considered their existence as an obvious fact, not deserving the
proof given by a real existence theorem. Even today the
overwhelming majority of the experimentalists know only Einstein's
old papers \cite{1}. The superficial (and \emph{false}) analogy
with Maxwell e.m. theory has generated ruinous consequences. Thus,
since forty years the GW hunters persist in their vain efforts
aimed at the experimental detection of GW's. In the last years the
inconclusive conclusions of their papers have assumed a pathetic
character of the following kind: -- No gravitational wave was
observed; however, the search provided us with an encouraging
upper limit on the gravity wave strain at our apparatuses. --

\par $\beta$) I reproduce here the abstract and the third paragraph
of the Introduction of a recent paper by S.D. Mohanty, Sz.
M\'{a}rka, \emph{et alii}, entitled ``Search algorithm for a
gravitational wave signal in association with Gamma Ray Burst
GRB030329 using the LIGO detectors'' \cite{7}. The abstract is as
follows: ``One of the brightest Gamma Ray Burst ever recorded,
GRB030329, occurred $[$on 29 March 2003$]$ during the second
science run of the LIGO detectors. At that time, both
interferometers at the Hanford, WA LIGO site were in lock and
acquiring data. The data collected from the two Hanford detectors
was $[$\emph{sic}!$]$ analyzed for the presence of a gravitational
wave signal associated with this GRB. This paper presents a
detailed description of the search algorithm implemented in the
current analysis.'' And the third paragraph of the Introduction
tells us that: ``The search algorithm is the component which acts
on pre-processed data and produces a list of candidate events or
the value of a statistics which is then used in drawing a
statistical inference. A first version of a full analysis pipeline
was reported in Mohanty S.D. \emph{et al.} 2004 \emph{Class.
Quantum Grav.} \textbf{21} S765-S774 and the pipeline used for the
present analysis will be described in LIGO Science Collaboration:
Abbott B. \emph{et al.} 2004 \emph{To be submitted}.''

\par Now, a report on INTERNET (13 January 2004) by the second
author of the above paper (Sz. M\'{a}rka) has informed us that
``We did not observe a gravity burst, which can be associated with
GRB030329.'' Any comment is superfluous. --

\vskip0.50cm

\begin{center}
$^{\star---------------------------\star}$
\end{center}

\end{document}